\documentstyle[preprint,aps,tighten,epsf]{revtex}

\draft
\begin{document}

\begin{center}
{\Large\bf Are citations of scientific papers a case of nonextensivity?}\\[15mm]

{\large Constantino Tsallis and Marcio P. de Albuquerque}\\[10mm]

Centro Brasileiro de Pesquisas Fisicas, Rua Xavier Sigaud 150 \\
22290-180 Rio de Janeiro-RJ, Brazil \\
E-mails: tsallis@cat.cbpf.br, mpa@cbpf.br\\[5mm]
{\em\today}
\end{center}


\begin{abstract}
\noindent
\end{abstract}
The distribution $N(x)$ of citations of scientific papers has recently been illustrated (on ISI and PRE data 
sets) and analyzed by Redner [Eur. Phys. J. B {\bf 4}, 131 (1998)]. To fit the data, a stretched 
exponential ($N(x) \propto \exp{-(x/x_0)^{\beta}}$) has been used with only partial success. The success 
is not complete because the data exhibit, for large citation count $x$, a power law (roughly $N(x) \propto 
x^{-3}$ for the ISI data), which, clearly, the stretched exponential does not reproduce. This fact is then 
attributed to a possibly different nature of rarely cited and largely cited papers. We show here that, within 
a nonextensive thermostatistical formalism, the same data can be quite satisfactorily fitted with a {\it 
single} curve (namely, $N(x) \propto 1/[1+(q-1)\; \lambda\; x]^{\frac{q}{q-1}}$ for the available values 
of $x$. This is consistent with the connection recently established by Denisov [Phys. Lett. A {\bf 235}, 
447 (1997)] between this nonextensive formalism and the Zipf-Mandelbrot law. What the present 
analysis ultimately suggests is that, in contrast to Redner's conclusion, the phenomenon might essentially 
be {\it one and the same} along the {\it entire} range of the citation number $x$.


\vspace{2cm}

Half a century ago, Zipf\cite{zipf} made his remarkable observations about some basic linguistic laws. 
More precisely, if we order the words appearing in a text (e.g., Homer's Iliad) from the most to the less 
frequent ones, thus obtaining a ranking (low rank for the most used, and high rank for the less used), we 
can plot, as a function of the rank, the number of times those words appear.  Zipf showed that, excepting 
the words with extremely low rank, an inverse power law emerges (so called {\it Zipf's law}). The 
exponent exhibits interesting universal aspects. For instance, for the spoken language, it appears to be 
very sensitive to the degree of instruction (primary, intermediate, highly academic) of the speaker, but 
very little to the particular culture (French, German, Anglo-saxon). Later on, Mandelbrot pointed out 
connections of this phenomenon with fractals\cite{mandelbrot}, and also suggested a further correction, 
namely that substantially better fittings can be obtained by using an inverse power law of the sum of the 
rank with a constant (so called {\it Zipf-Mandelbrot law}). A further step along this line was provided 
recently by Denisov\cite{denisov}. Indeed, using within the Sinai-Bowen-Ruelle thermodynamical 
formalism for symbolic dynamics, the nonextensive thermostatistics proposed some years ago by one of 
us\cite{tsallis}, Denisov {\it deduced} the Zipf-Mandelbrot law. To be more precise, it is clear that, 
unless one uses a specific model, there is no way to deduce the {\it precise values} for the exponent and 
the additive constant. What Denisov deduced, from very generic entropic arguments, was the {\it form} 
of the law. In this sense, the approach is very analogous to those which succeed associating Gaussians to 
normal diffusion, and Levy or Student's $t$-distributions to anomalous superdiffusion (see, for instance, 
\cite{tsallislevy} and \cite{tsallisbukman} respectively). Finally, it is important to stress here that, 
although the present problematic was historically triggered in Linguistics, the same kind of 
considerations are equally relevant to DNA sequences, artificial languages, and a variety of  other 
stochastic, deterministic or mixed processes.

Here we focus on an interesting analysis of data concerning citations of scientific publications. More 
precisely, Redner\cite{redner} recently exhibited and discussed the distributions of citations related to 
two quite large data sets, namely (i) 6 716 198 citations of  783 339 papers, published in 1981 and cited 
between 1981 and June 1997, that have been catalogued by the Institute for Scientific Information (ISI), 
and (ii) 351 872 citations, as of June 1997, of 24 296 papers cited at least once and which were published 
in Physical Review D (PRD) in volumes 11 through 50 (1975-1994). In his study, Redner addressed the 
citations of publications, in variance with Laherrere and Sornette\cite{sornette}, who adressed, in a 
similar study, the citations of authors. If we denote by $x$ the number of citations and by $N(x)$ the 
number of papers that are cited $x$ times. The main results of the study were that, for relatively large 
values of $x$, $N(x) \propto 1/x^{\alpha}$ with 
$\alpha \simeq 3$, whereas, for relatively small values of $x$, the data were reasonably well fitted with a 
stretched exponential, i.e., $N(x) \propto \exp [-(x/x_0)^{\beta}]$, $\beta$ and  $x_0$ being the fitting 
parameters ($\beta \simeq 0.44$ and  $0.39$ for the ISI and the PRD data respectively); see Figs. (a) and 
(b). Since a streched exponential by no means asymptotically provides an inverse power law, the author 
concluded that large $x$ and low $x$ phenomena are of {\it different nature} (in the author's words, " 
These results provide evidence that the citation distribution is not described by a single function over the 
entire range of citation count"). While the phenomenon exhibited by Redner is of great interest, we must 
disagree with his conclusion. It is the central purpose of our present effort to develop arguments within 
the nonextensive statistical mechanics mentioned above\cite{tsallis}, and along the lines of Denisov, 
which will lead to a {\it single} function $N(x)$ having, like the streched exponential, only two fitting 
parameters. This function is of the power-law type and will turn out to fit both ISI and PRD experimental 
data sensibly better than the forms described above.

Before presenting our arguments for this specific problem, let us briefly introduce the nonextensive 
formalism we are referring to.  If the physical system we are focusing on involves long-range interactions 
or long-range microscopic memory or (multi)fractal boundary conditions, it can exhibit a quite 
anomalous thermodynamic behavior, which might even be untractable within Boltzmann-Gibbs (BG) 
statistical mechanics. To overcome at least some of these pathological situations, an entropic form $S_q$ 
has been proposed\cite{tsallis} which yields a generalization of standard statistical mechanics and 
thermodynamics. This entropy is defined as follows:
\begin{equation} 
S_q \equiv k\frac{1-\sum_ip_i^q}{q-1}\;\;\;\;\;\; (\sum_ip_i=1;\; q \in \cal{R})
\end{equation}
where $k$ is a positive constant (from now on taken to be unity, without loss of generality). In the limit 
$q \rightarrow 1$, we recover the usual BG entropy, i.e., $S_1=-\sum_ip_i\;\ln p_i$. A property which 
characterizes the above generalized entropic form is the following: if we have two independent systems 
$A$ and $B$ such that $p_{ij}^{A+B}=p_i^A \;p_j^B$, then
\begin{equation}
S_q(A+B)=S_q(A)+S_q(B)+(1-q)S_q(A)S_q(B)
\end{equation}
Consequently, if $q>1,=1$ or $<1$, $S_q$ is {\it subextensive, extensive or superextensive}. 
Optimization of this entropy with appropriate constraints provides equilibrium distributions which are of 
the power-law type, and which recover the exponential Boltzmann factor only in the limit $q \rightarrow 
1$. 

This thermostatistics has provided interesting insights onto a variety of physical
systems such as two-dimensional turbulence in pure-electron plasma\cite{boghosian}, self-gravitating 
systems\cite{plastino}, cosmology\cite{cosmology}, solar neutrinos\cite{neutrino}, 
Levy\cite{tsallislevy} and correlated\cite{tsallisbukman} anomalous diffusions,
inverse bremsstrahlung absorption in plasma\cite{andre}, quantum scattering\cite{ion}, one-dimensional 
maps\cite{maps}, a variety of self-organized
critical models\cite{SOC}, long-range interaction conservative systems\cite{celia}, processing of EEG 
signals of epileptic humans and turtles\cite{EEG}, among others (see \cite{tsallisbjp} for a review). To 
theoretically study such complex systems within this nonextensive formalism, some approaches (besides,
naturally, the usual analytic and numerical methods) are now available such as the generalizations
of (i) Kubo's linear response theory, (ii) Feynman's perturbation theory as well as the
Bogoliubov's inequality (basis of the variational method), (iii) Green's functions, and (iv) 
Feynman's path integral (respectively generalized, in the realm of nonextensivity, in 
\cite{rajagopal1,lenzi,rajagopal2,renio}).  

Let us now focuse on our specific problem, namely the distributions of citations. The corresponding 
entropic form is given by
\begin{equation}
S_q=\frac{1-\sum_{x=1}^{\infty}p_x^q }{q-1} 
\end{equation}
The optimization of this entropy with the corresponding constraints\cite{tsamepla}, namely
\begin{equation}
\sum_{x=1}^{\infty}p_x=1
\end{equation}
and
\begin{equation}
<x>_q \;\equiv\frac{\sum_{x=1}^{\infty}x\;p_x^q}{\sum_{x=1}^{\infty}p_x^q}=constant\;\;,
\end{equation}
yields
\begin{equation}
p_q(x)=\frac{[1-(1-q)\lambda x]^{\frac{1}{1-q}}}{\sum_{y=1}^{\infty} [1-(1-q)\lambda 
y]^{\frac{1}{1-q}}  }
\end{equation}
where, unless $q=1$, $\lambda$ is {\it not} (see \cite{tsamepla}) the Lagrange parameter associated with 
constraint (5), but can nevertheless be determined through that constraint. This distribution is expected to 
be an excellent approximation for $x$ not too small (say, not below 5), but departures would not be 
surprising while approaching unity. Indeed, we have deduced Eq. (6) through generic entropic 
considerations and not by using a specific model. Also, for precisely the same reason, $q$ and $\lambda$ 
are to be considered as free parameters within the present phenomenological approach. 

Eq. (6) implies that the so-called {\it escort} distribution is given by
\begin{equation}
P_q(x) \equiv \frac{[p_q(x)]^q}{\sum_{y=1}^{\infty}[p_q(y)]^q}=
\frac{[1-(1-q)\lambda x]^{\frac{q}{1-q}}}{\sum_{y=1}^{\infty} [1-(1-q)\lambda y]^{\frac{q}{1-q}}  }
=\frac{ \Big\{  \sum_{y=1}^{\infty}\frac{1} {[1+(q-1)\lambda y]^{\frac{q}{q-1}}}\Big\}^{-1}   
}{[1+(q-1)\lambda x]^{\frac{q}{q-1}}}
\end{equation}
This escort distribution is to be identified (see \cite{tsamepla}) with the above introduced experimental 
distribution $N(x)$, hence
\begin{equation}
N(x)=N(1)\frac{[1+(q-1)\lambda ]^{\frac{q}{q-1}} }{[1+(q-1)\lambda x]^{\frac{q}{q-1}}}
\end{equation}
or, equivalently,
\begin{equation}
N(x)=\frac{N_0 }{[1+(q-1)\lambda x]^{\frac{q}{q-1}}}
\end{equation}
where we have simplified the notation by introducing $N_0$. The fittings of both ISI and PRD data series 
using this functional form are exhibited in Figs. (c) and (d). We can appreciate that they are considerably 
better (in both precision and completeness) than those appearing in \cite{redner}. In particular, we have 
obtained, for the ISI series, $q\simeq1.53$, hence $q/(q-1)\simeq2.89$, which is clearly compatible with 
the approximate exponent 3 advanced in \cite{redner}.

As a summarizing conclusion, we suggest that, in variance with what is stated in \cite{redner}, the 
present interesting linguistic-like phenomenon revealed by Redner appears to emerge from one and the 
same reason for practically the {\it entire} range of citation score $x$. Furthermore, this reason appears 
to be deeply related to thermostatistical nonextensivity. Specific microscopic models are of course very 
welcome in order to achieve a more concrete insight, and also for addressing the exceedingly small 
values of $x$, which are out of the scope of the present phenomenological approach.

Finally, CNPq and PRONEX (Brazilian Agencies) are acknowledged for partial support.

\newpage

{\bf Caption:} ISI and PRD distributions of citations (experimental data and fittings). From \cite{redner}: 
(a) log-linear plot and (b) log-log plot. Present work: (c) log-linear plot and (d) log-log plot (Eq.(9) has 
been used with  the values for $(q, \lambda, N_0)$ indicated in the figure).


\begin{thebibliography}{99}

\bibitem{zipf}G.K. Zipf, {\it Human behavior and the principle of least effort} 
(Addison-Wesley, Cambridge-MA, 1949).

\bibitem{mandelbrot}B.B. Mandelbrot, {\it The fractal geometry of nature} 
(Freeman, San Francisco, 1983).

\bibitem{denisov}S. Denisov, Phys. Lett. A {\bf 235}, 447 (1997).

\bibitem{tsallis} C. Tsallis, J. Stat. Phys. {\bf{52}}, 479 (1988); E.M. F. Curado and C. Tsallis, J. Phys. 
A: Math. Gen. {\bf{24}}, L69 (1991) [Corrigenda: {\bf{24}}, 3187 (1991); {\bf 25}, 539 (1992)]. A 
regularly updated bibliography is available at http://tsallis.cat.cbpf.br/biblio.htm

\bibitem{tsallislevy} D.H. Zanette and P.A. Alemany, Phys. Rev. Lett. {\bf 75}, 366 (1995); C. Tsallis,
S.V.F. Levy, A.M.C. de Souza and R. Maynard, Phys. Rev. Lett. {\bf 77}, 5442 (1996); {\bf 77},
5442(E) (1996); C. Tsallis and D. Prato, {\it Revisiting the nonextensive foundation of Levy 
distributions}, preprint (1999).

\bibitem{tsallisbukman}A.R. Plastino and A. Plastino, Physica A {\bf 222}, 347 (1995); C. Tsallis and 
D.J. Bukman, Phys. Rev E {\bf 54}, R2197 (1996); A.M.C. Souza and C. Tsallis, Physica A  {\bf 236}, 
52 (1997).

\bibitem{redner}S. Redner, Eur. Phys. J. B {\bf 4}, 131 (1998).

\bibitem{sornette}J. Laherrere and D. Sornette, Eur. Phys. J. B {\bf 2}, 525 (1998).

\bibitem{boghosian}B.M. Boghosian, Phys. Rev. E {\bf 53}, 4754 (1996).

\bibitem{plastino}A.R. Plastino and A. Plastino, Phys. Lett. A {\bf 174}, 384 (1993) and {\bf 193}, 251 
(1994).

\bibitem{cosmology} V.H. Hamity and D.E. Barraco, Phys. Rev. Lett. {\bf 76}, 4664 (1996); D.F. 
Torres, H. Vucetich and A. Plastino, Phys. Rev. Lett. {\bf 79}, 1588 (1997) and {\bf 80}, 3889(E) (1998). 

\bibitem{neutrino}G. Kaniadakis, A. Lavagno and P. Quarati, Phys. Lett. B {\bf 369}, 308 (1996); P. 
Quarati et al, Nucl. Phys. A {\bf 621}, 345c (1997).

\bibitem{andre}C. Tsallis and A.M.C. de Souza, Phys. Lett. A {\bf 235}, 444 (1997).

\bibitem{ion}D.B. Ion and M.L.D. Ion, Phys. Rev. Lett. {\bf 81}, 5714 (1998).

\bibitem{maps}C. Tsallis, A.R. Plastino and W.-M. Zheng, Chaos, Solitons and Fractals {\bf 8}, 885
(1997); M.L. Lyra and C. Tsallis, Phys. Rev. Lett. {\bf 80}, 53 (1998); U. Tirnakli, C. Tsallis and M.L. 
Lyra, Eur. Phys. J. B (1999), in press.

\bibitem{SOC}F.A. Tamarit, S.A. Cannas and C. Tsallis, Eur. Phys. J. B {\bf 1}, 545 (1998); A.R.R. 
Papa and C. Tsallis, Phys. Rev. E {\bf 57}, 3923 (1998).

\bibitem{celia}C. Anteneodo and C. Tsallis, Phys. Rev. Lett. {\bf 80}, 5313 (1998).

\bibitem{EEG}L.G. Gamero, A. Plastino and M.E. Torres, Physica A {\bf 246}, 487 (1997); A. Capurro, 
L. Diambra, D. Lorenzo, O. Macadar, M.T. Martin, C. Mostaccio, A. Plastino, J. Perez, E. Rofman, M.E. 
Torres and J. Velluti, Physica A {\bf 265}, 235 (1999).

\bibitem{tsallisbjp}C. Tsallis, in {\it Nonextensive Statistical Mechanics and Thermodynamics}, eds. 
S.R.A. Salinas and C. Tsallis, Braz. J. Phys. {\bf 29}, 1 (1999).[cond-mat/9903356].

\bibitem{rajagopal1}A.K. Rajagopal, Phys. Rev. Lett. {\bf 76}, 3469 (1996).

\bibitem{lenzi}E.K. Lenzi, L.C. Malacarne and R.S. Mendes, Phys. Rev. Lett. {\bf 80}, 218 (1998).

\bibitem{rajagopal2}A.K. Rajagopal, R.S. Mendes and E.K. Lenzi, Phys. Rev. Lett. {\bf 80}, 3907
(1998).

\bibitem{renio}E.K. Lenzi, L.C. Malacarne and R.S. Mendes, preprint (1998).

\bibitem{tsamepla}C. Tsallis, R.S. Mendes and A.R Plastino, Physica A {\bf 261}, 534 (1998).
\end{thebibliography}
\end{document}